\documentstyle[12pt]{article}
\newcommand{\bb}{\begin{eqnarray}}
\newcommand{\ee}{\end{eqnarray}}
\begin{document}
\title{{Entropy of extremal black holes}}
\author{P. Mitra\thanks{e-mail mitra@tnp.saha.ernet.in}\\
Saha Institute of Nuclear Physics\\
Block AF, Bidhannagar\\
Calcutta 700 064, INDIA}
\date{hep-th/9704201}
\maketitle
\begin{abstract}
After summarizing the development of black hole thermodynamics in
the seventies, we describe a recent microscopic model. This model
suggests that  the  Bekenstein-Hawking  area  formula  holds  for
extremal black holes as well as for ordinary (non-extremal) ones.
On the other hand, semiclassical studies have suggested a discontinuity
between  non-extremal  and  extremal  cases. We indicate how a
reconciliation has been brought about by summing over topologies.
\end{abstract}
\bigskip
\begin{center}
{\it Based on Invited Talk at
Workshop  on  Frontiers  of  Quantum Field  Theory,  Quantum
Gravity  and  Strings: Puri, December 1996}
\end{center}
\bigskip
\section{Introduction}
In Einstein's theory of gravitation, the gravitational field due
to a point mass is described by a metric which has many interesting
properties. Its black hole features have been known for a very
long time, but in the seventies it began to appear that thermodynamic
concepts like temperature and entropy were also associated with it.
Gradually it was realized that these were quantum effects. But the
degrees of freedom associated with the entropy could not be clearly identified.
Many suggestions have been made. A recent one
made in 1996 itself involves the embedding of some black holes
in string theory. It has been possible to identify the quantum states
contributing to the black hole entropy, which naturally has the standard value
\cite{vafa}. However, these embeddings
involve supersymmetry and are not quite universal. One would like to
identify the relevant states in the original theory instead of the embedded
one. It is to be hoped that the coming years will bring further progress.

Meanwhile, in this talk a more conventional framework will be used to
discuss the entropy of what are known as {\it extremal} black
holes. Over the past couple of years it has been unclear whether the
so-called Bekenstein-Hawking formula is applicable to these black
holes. There is no evidence that the originators of this formula supposed
it to continue to hold for the special class; indeed, some of the
arguments which were originally used to derive this formula indicate
that the formula may fail in the extremal case. On the other hand several
non-critical authors have used the formula loosely without pausing to
think whether all black holes have to obey it.  Fortunately, a clear
picture does seem to have emerged on this issue now, as we hope to
explain.

After summarizing the history of black hole entropy as it developed in
the seventies, we first refer to a model which may be said to have
indicated that extremal black holes are related to other black holes
in a not particularly discontinuous way (this model came
after the string theory developments mentioned above.) Thereafter
we go back to the usual semiclassical approach to extremal black holes
and recall the glaring indications of discontinuity in that approach. 
As we understand the situation now, this discontinuity arises in {\it one}
way of quantization of the classical theory. An alternative way which
leads to the Bekenstein-Hawking formula even for extremal black holes
is reviewed in some detail.
\section{Black hole entropy in the seventies}
A precursor of the idea of entropy in the context of black holes
was the so-called area theorem \cite{Hawk71}. According to this theorem,
the area of the horizon of a system of black holes always increases
in a class of spacetimes. The asymmetry in time is built into the
definition of this class: these spacetimes are predictable from
partial Cauchy hypersurfaces. This result is certainly reminiscent
of thermodynamical entropy.

Some other observations made around that time were collected together
into a set of {\it laws of black hole mechanics} analogous to the
laws of thermodynamics \cite{BCH}.\begin{itemize}
\item The zeroeth law states that the surface gravity $\kappa$
remains constant on the horizon of a black hole.
\item The first law states that
\bb
{\kappa dA\over 8\pi}=dM-\phi dQ,
\ee
where $A$ represents the area of the horizon and $\phi$ the potential
at the horizon.
For the Reissner - Nordstr\"{o}m black hole, with horizons at
\bb
r_{\pm}=M\pm\sqrt{M^2-Q^2},
\ee
\bb
\kappa={r_+-r_-\over 2r_+^2},~ \phi={Q/r_+},~ A=4\pi r_+^2.
\ee
\item The second law is just the area theorem already stated.
\end{itemize}.

When   these   observations  were  made,  there  was  no  obvious
connection with thermodynamics, it was only a matter of  analogy.
But  it  was  soon  realized  \cite{Bek}  that the existence of a
horizon  imposes  a  limitation  on  the  amount  of  information
available  and hence may lead to an entropy, which should then be
measured by the geometric quantity associated with  the  horizon,
namely its area.
Thus,  upto  a  factor,  $A$  should represent the entropy and ${
\kappa\over 8\pi}$ the temperature.

This interpretation of the  laws  of  black
hole  mechanics was not fully convincing,
and  in any case the undetermined factor left a
question mark. Fortunately, the problem was solved very soon.  It
was discovered that quantum theory causes dramatic changes in the
behaviour of black hole spacetimes. A scalar field theory in  the
background of a Schwarzschild black hole indicates the occurrence
of radiation of particles \cite{Hawk} at a temperature
\bb
T={\hbar\over 8\pi M}={\hbar\kappa\over 2\pi }.\label{T}
\ee
This demonstrated the  connection  of  the  laws  of  black  hole
mechanics  with  thermodynamics  and  fixed  the scale factor. It
involves Planck's constant and is a quantum effect.

For  a  Schwarzschild  black  hole,  the  first   law   of   {\it
thermodynamics} can be written as
\bb
TdS=dM
\ee
and can be integrated, because of (\ref{T}), to yield
\bb
S={4\pi M^2\over \hbar}={A\over 4 \hbar}.
\ee

Although  the  expression  for $T$ given above is specific to the
case of Schwarz\-schild  black  holes,  the  relation  between  the
temperature  and  the  surface gravity given in (\ref{T}) is more
generally valid in the case of black holes having $g_{tt}\sim (1-
{r_h\over r})$. The first law of black hole mechanics then becomes
\bb
T d{A\over 4\hbar}=dM-\phi dQ.
\ee
Comparison with the first law of thermodynamics
\bb
{T dS}=dM-\tilde\phi dQ
\ee
is not straightforward because  the  chemical  potential  $\tilde
\phi$  is not clearly known. However, one way of satisfying these
two equations involves the identification
\bb
S={A\over 4\hbar},\qquad \tilde\phi=\phi.\label{std}
\ee

In another approach, the grand partition function is used. For
charged black holes \cite{GH} it
can be related to the classical action by
\begin{equation}
Z_{\rm grand}=e^{-{M- TS-\tilde\phi Q\over T}}\approx e^{-I/\hbar},
\end{equation}
where  the  functional  integral  over  all   configurations  
consistent with the appropriate boundary conditions is 
semiclassically approximated  by  the  exponential weight factor for the
classical action $I$ of the black hole. This action (see below) is given
by a quarter of the area of the horizon when the Euclidean time goes
over one period, {\it
i.e.,} from zero to $\hbar/T$. Consequently,
\bb
M=T(S+{A\over 4\hbar})+\tilde\phi Q.\label{semi}
\ee
Now  there  is a standard formula named after Smarr \cite{Smarr},
\bb
M={\kappa A\over 4\pi}+\phi Q,
\ee
which can be rewritten as
\bb
M=T{A\over 2\hbar}+\phi Q.\label{smarr}
\ee
Comparison with (\ref{semi}) suggests once  again  the  relations
(\ref{std}).  Although the result is the same, it should be noted
that there is a new input: the functional integral.  There  is  a
hope  that  corrections  to the above formulas may be obtained by
improving the  approximation  used  in  the  calculation  of  the
functional integral.

\subsection{On-shell action}
To  see  that  the  action  equals  a quarter of the area, let us
consider a euclidean Reissner - Nordstr\"{o}m  black  hole  in  a  manifold
$\cal M$ with a boundary which is subsequently taken to infinity.
The action has the expression
\bb
I&=&-{1\over  16\pi}\int_{\cal M} d^4x\sqrt gR+{1\over 8\pi}\int_{\partial
{\cal M}} d^3x\sqrt\gamma (K-K_0)+\nonumber\\ &&{1\over 16\pi}\int_{\cal M}
d^4x\sqrt g F_{\mu\nu} F^{\mu\nu}.\label{action}
\ee
Here  $\gamma$ is the induced metric on the boundary $\partial {\cal M}$
and  $K$ the extrinsic curvature, from which a subtraction has to
be made to make the action finite.

The  first  term  of  the  action  vanishes  because   Einstein's
equations lead to $R=0$.

To evaluate the second term, we take the boundary of the manifold
at $r=r_B\to\infty$. Then
\bb
K&=&-{1\over\sqrt{g_{tt}}r^2}{1\over\sqrt{g_{rr}}}{d\over dr}
(\sqrt{g_{tt}}r^2)\nonumber\\ &=&-{1\over r^2}{d\over dr}
[(1-{M\over r}+\cdots)r^2]\nonumber\\ &=&-{1\over r^2}{d\over dr}
(r^2-Mr),
\ee
and
\bb
\int d^3x\sqrt\gamma=\int dt(1-{M\over r}+\cdots)4\pi r^2.
\ee
We  see  that  $\int d^3x\sqrt\gamma K$ diverges as $r\to\infty$,
but this can be cured by subtracting  from  $K$  the  flat  space
contribution  $K_0=-{1\over r^2}{d\over dr}r^2$. The second piece
of the action becomes
\bb
&&-{1\over 8\pi}\int dt(1-{M\over r}+\cdots)4\pi r^2 {1\over r^2}
{d\over dr}(-Mr)|_{r=r_B\to\infty} \nonumber\\ &=&-{1\over 2}\int dt (-M)
={1\over 2}\beta M.
\ee

Finally, the third term of the action becomes
\bb
&&-{1\over 16\pi}\int dt .4\pi\int d r^2. 2.{Q^2\over
r^4}\nonumber\\ &=&-{1\over 2}\int dt{Q^2\over r_+} \nonumber\\ &=&
-{1\over 2}\beta Q\phi,
\ee
where  $\phi$ is the electrostatic potential at the horizon. The
sign is negative here  because  in  the  euclidean  solution  the
electric field is purely imaginary.

Putting  all  pieces of the action together, we find the numerical value
of the action to be
\bb
I={1\over 2}\beta(M-Q\phi)={A\over 4}.
\ee
As indicated above, this leads to an entropy of the same value (in
natural units).

\section{Microscopic model for near-extremal black holes}
It has recently been observed (cf. \cite{alwis}) that a one-dimensional
gas of massless particles
can be used as a model for  black  holes  in  any  number  of
dimensions.   The   particles   can   be  either  left-moving  or
right-moving -- there is no mixing between the two types. Both bosons
and fermions  can  be  present.  If  the  total  length  of  the
one-dimensional  space is $L$, the entropy and the energy are given
by
\bb
S={\pi L\over 6\hbar}[n_LT_L+n_RT_R],~
E={\pi L\over 12\hbar}[n_LT_L^2+n_RT_R^2],
\ee
where $n_L(n_R)$ is the number of left(right)-moving bosons  plus
half  the  corresponding  number  of  fermions. In the absence of
interactions,  the  left  and  right  degrees  of  freedom   are
independent, and the corresponding temperatures can be different.
The  effective  temperature  may  be   defined   by   $({\partial
S\over\partial E})^{-1}$, the differentiation being carried out at
constant momentum, {\it i.e.,} constant difference between $E_L$
and $E_R$. This leads to a temperature
\bb
T={2T_LT_R\over T_L+T_R}
\ee
equal to the harmonic mean of $T_L$ and $T_R$.
If $n_L=n_R=n$, these equations get somewhat simplified and one has
\bb
E={\pi nL\over 12\hbar}[({6\hbar S\over \pi nL})^2-
{6\hbar ST\over \pi nL}],
\ee
whence,
\bb
{12\hbar S\over \pi nL}=T+\sqrt{T^2+{48E\over\pi nL}}.
\ee
To compare these quantities with those for a near-extremal charged black hole
in four dimensions, we put
\bb
E=E_0+\epsilon,~T=T(M,Q)=T(Q+\epsilon,Q)
\ee
to get
\bb
S=\sqrt{\pi nLE_0\over 3}+{nL\over 24}\sqrt{2\epsilon\over Q^3}
+\epsilon[\sqrt{\pi nL\over 12E_0} -{ nL\over 6Q^2 }+
{(\pi nL)^{3/2}\over 192\sqrt{3}\pi^2Q^3\sqrt{E_0}}]+\cdots.
\ee
Comparison with the area formula
\bb
S=\pi(Q^2+2Q\sqrt{2Q\epsilon}+2Q\epsilon)+\cdots
\ee
for the black hole shows that agreement occurs for a continuous range
of values of $\epsilon$ provided that
\bb
nL=48\pi Q^3,~~E_0={Q\over 16}.
\ee
The first equality here relates the number $n$ to the
parameter $Q$ characterizing the family of
black holes being considered; the second equality fixes a zero-point
shift. If  these conditions are satisfied, the gas of massless particles
can  be  regarded  as  a  model   for   the family of near-extremal
black   holes.   The
one-dimensional  particles  can  be  modes  of  a string. In this
sense, the model may be embedded into string theory \cite{alwis}.
In any case, the model indicates the entropy of the black hole family
to be continuous in the limit $\epsilon\to 0$.

\section{Extremal black holes}
In the recent past there has been special interest in extremal
black holes. First it was pointed out \cite{GMone} that the entanglement
entropy, which is usually proportional to the area of a black hole,
ceases to be so for extremal black holes. Thereafter, \cite{HHR}
noticed that euclidean topology is discontinuous in the passage from
non-extremal to extremal black holes and argued that the entropy of
extremal black holes might actually vanish. In \cite{GMtwo} it was
shown that this argument could be relaxed and an extremal black hole
could be allowed to have an entropy proportional to the mass. Model
calculations analogous to \cite{alwis} developed formulations
that are continuous in the limit of non-extremal black holes
going into extremal ones. There is then something of a contradiction.
Is there a discontinuity, or is there none?

\subsection{Vanishing Action}
To  see  that  the  action  vanishes in the extremal case, and is thus
discontinuous, let us once again
consider the euclidean Reissner - Nordstr\"{o}m  black  hole  in the manifold
with  boundary.

The  first  term  of  the  action (\ref{action})  vanishes again because
Einstein's equations lead to $R=0$.
As before, the second piece of the action is
${1\over 2}\beta M$.
The third term of the action is
\bb
&&-{1\over 16\pi}\int dt. 4\pi\int dr^2 .2.{Q^2\over
r^4}\nonumber\\ &=&-{1\over 2}\int dt{Q^2\over r_+} \nonumber\\ &=&
-{1\over 2}{\beta M^2\over r_+}\nonumber\\ &=&
-{1\over 2}\beta M
\ee
as $Q=M=r_+$ here.

Putting  all  pieces of the action together, we find the value of
the action to be
\bb
I={1\over 2}\beta(M-M)=0.
\ee
In doing this calculation, $\beta$ has been assumed finite. If the
extremal limit of a non-extremal black hole is taken, this quantity
actually goes to infinity. However, as pointed out in \cite{HHR},
there is no conical singularity in the extremal case, so that there is
no reason to fix the temperature in this case, and the temperature should be
regarded as arbitrary.

\subsection{Entropy proportional to mass}
The laws of black hole physics  suggest  that  nonextremal  black
holes possess an entropy proportional to the area of the
horizon. When the scale
is fixed by comparing the temperature thus suggested with that given by
the semiclassical calculations of \cite{Hawk}, the entropy turns out to be a
quarter of the area. If one is interested in an extremal black hole, one
may be tempted to regard it as a special limiting case of a sequence
of nonextremal black holes and thus infer that the same formula should
hold for the entropy. But it was pointed out in the context of Reissner -
Nordstr\"{o}m black holes \cite{HHR} that the
extremal and nonextremal cases of
the euclidean version are topologically different, so that continuity
need not hold.
Moreover, it was shown in \cite{GMtwo} that
the  derivation  of  an  expression for the thermodynamic entropy
of an extremal black hole following \cite{GH}
allows an extra term proportional to the mass
of the black hole.
It will be instructive to elaborate a little on
the discussion of the Reissner - Nordstr\"{o}m black hole in \cite{GMtwo}.

For a charged black hole, the first law of thermodynamics
\begin{eqnarray}
TdS=dM-\Phi dQ,\end{eqnarray}
involves two ``intensive" variables, {\it viz.} $T$,  the temperature,
conjugate to $M$, and $\Phi$ the chemical potential, conjugate to $Q$.
We are interested in the extremal case $Q=M$. Then there is only one
independent thermodynamical variable, $Q$ or $M$, so the first law
should involve only one conjugate variable and can be written as
\begin{eqnarray}
dS=\gamma dM.\label{xfl}\end{eqnarray}
If this equation is sought to be understood in terms of the previous one,
$\gamma$ must be interpreted as ${1-\Phi\over T}$ (see below).

To understand the meaning of $\gamma$, one
has to imagine a thermodynamic system of mass $M$ and charge $Q$
in contact with a reservoir of energy and charge such that exchanges
of energy and charge with the system are always constrained to be equal.
In this situation, the total change of entropy of the system and the
reservoir is given by
\begin{eqnarray}
dS_{tot}=\gamma dM-{dM\over T_{reservoir}}+{\Phi_{reservoir}dM
\over T_{reservoir}}.
\end{eqnarray}
The condition for equilibrium is then
\begin{eqnarray}
\gamma ={1-\Phi_{reservoir} \over T_{reservoir}}.
\end{eqnarray}
Thus, instead of the usual equality of temperatures and chemical
potentials, there is only {\it one} condition, with $\gamma$ equalling
a certain combination of the temperature and the chemical potential
of the reservoir. In other words, one cannot even talk separately
of a temperature and a chemical potential for the system: there is only
this combination $\gamma$. Correspondingly, the ensemble is not quite
grand canonical, but a {\it reduced} grand canonical one.

Indeed, the partition function also has to be written as
\begin{eqnarray}
Z=e^{S-\gamma M}=e^{-I}.\end{eqnarray}
Here, $I$ is the effective action, which is set equal to the
classical on-shell action in the lowest approximation, and
vanishes in the extremal case, as seen above.  This implies
\begin{eqnarray}
S=\gamma M.\end{eqnarray}
Comparison with the first law (\ref{xfl}) then shows that
\begin{eqnarray}
d\gamma=0,\end{eqnarray}
so that $\gamma$ is a constant, hence
the entropy is a constant times the mass.
This constant may of course vanish, but that is a special case.

\subsection{The area law again}
While the possibility of microscopic models is interesting, the suggestion
that the entropy is continuous in the extremal limit
is intriguing in view of the developing belief that the area formula
applies only to non-extremal black holes. It is true that
the borderline between non-extremal and extremal cases is very thin and
if one takes the extremal limit of non-extremal black holes instead of
an extremal black hole directly, one obtains the area answer. But, as
mentioned above, the euclidean topologies are different, so one should
consider not the limit but the extreme black hole by itself; and then
the semiclassical approach does not yield the area law.
A simple way out of this mismatch would be to say that the model
is wrong, but it would certainly be more positive to look for a
way of obtaining the area answer directly for
an extremal black hole.

Usually, when one quantizes a classical theory, one tries to preserve
the classical topology. In this spirit, one usually seeks to have
a quantum theory of extremal black holes based exclusively on
extremal topologies. As an alternative, we shall
try out a quantization where a sum over topologies is carried out.
Thus, in our consideration of the functional integral, classical
configurations corresponding to both topologies will be included. The
extremality condition will then be imposed
on the averages that result from the
functional integration. We shall, following \cite{GH} and \cite{york},
use a grand canonical ensemble. Here the temperature and the chemical
potential are supposed to be specified as inputs, and the average mass $M$
and charge $Q$ of the black hole are outputs. So the actual definition
of extremality that we have in mind for a Reissner- Nordstr\"{o}m
black hole is $Q=M$. This may be described  as {\it extremalization after
quantization}, as opposed to the usual approach of {\it quantization after
extremalization.}\cite{GMthree}

The action for the euclidean version of a Reissner - Nordstr\"{o}m
black hole on a four dimensional manifold ${\cal M}$ with a boundary
has been given in (\ref{action}). A class of spherically symmetric
metrics \cite{york} is considered on ${\cal M}$:
\bb
ds^2=b^2d\tau^2+\alpha^2dr^2+r^2d\Omega^2,
\ee
with  the variable $r$ ranging between $r_+$ (the horizon) and $r_B$ (the
boundary), and $b, \alpha$ functions of $r$  only.  There  are
boundary conditions as usual:
\bb
b(r_+)=0,~2\pi b(r_B)=\beta. 
\ee
Here  $\beta$  is the inverse temperature and $r_B$ the radius of
the boundary which will be taken to infinity.
There is another boundary condition involving $b'(r_+)$:
It reflects the extremal/non-extremal  nature  of  the  black  hole  and  is
therefore   different   for  the  two conditions:
\bb
{b'(r_+)\over\alpha(r_+)}&=&1{\rm ...in~non-extremal~case},\nonumber\\
{\rm but~}&=&0 {\rm ...in~extremal~case}.
\ee

The  vector
potential is taken to be zero and the scalar potential  satisfies
the boundary conditions
\bb
A_\tau(r_+)=0, ~A_\tau(r_B)={\beta\Phi\over 2\pi i}.
\ee

The action (\ref{action}) with  this  form  of  the  metric  depends  on the
functions $ b(r), \alpha(r)$ and $A_\tau(r)$: this may be regarded as
a reduced action. Variation of these functions with proper
boundary  conditions  leads  to  reduced versions of
the  Einstein  -  Maxwell
equations. The solution of a subset of these equations,
namely the Gauss law and the Hamiltonian constraint, is given by \cite{york}
\bb
{1\over\alpha}=[1-{2m\over r}+{q^2\over r^2}]^{1/2}, ~A'_\tau=
-{iqb\alpha\over r^2},
\ee
with the mass parameter $m$ and the  charge $q$ arbitrary.
The reason why these parameters have not been
expressed as functions of $\beta,\Phi$ is that some of the equations
of motion and the corresponding boundary conditions
have not been imposed on the solution. Instead of that, the
action may be expressed in terms of $m,q$ and then extremized with respect to
$m,q$ \cite{york}.

The value of the action corresponding to the solution depends
on the boundary condition:
\bb
I&=& \beta(m-q\Phi) -\pi (m+\sqrt{m^2-q^2})^2
{\rm ~for~non-extremal~bc},\nonumber\\
I&=&\beta(m- q\Phi) {\rm ~for~extremal~bc}.\label{I}
\ee
The first line is taken from \cite{york},
where the non-extremal boundary condition was used
in connection with a semiclassically quantized  non-extremal  black
hole. The second line corresponds to  the extremal boundary condition
used in connection with  a  semiclassically
quantized   extremal   black   hole
\cite{GMcom}.  As the euclidean
topologies  of  non-extremal  and  extremal  black   holes   are
different, quantization was done separately for the two cases in
\cite{york,GMcom}. The topology was selected before quantization.

As indicated above, a different approach has to be used here.  The
two topologies are to be summed over in the functional integral \cite{GMthree}
and the extremality condition imposed afterwards.

Thus the partition function is of the form
\bb
\sum_{\rm topologies}\int d\mu(m)\int d\mu(q) e^{-I(q,m)},
\ee
with   $I$   given    by    (\ref{I})    as    appropriate    for
non-extremal/extremal $q$.

The  semiclassical approximation involves
replacing the double integral by the
maximum value of the integrand, {\it i.e.,} by the
exponential of the negative of the minimum  $I$.
We consider the variation of $I$ as $q,~m$ vary in both topologies.
It  is  clear from (\ref{I})
that the non-extremal action is lower than the extremal one for each set of
values of $q,~m$.
Consequently, the partition function is  to  be  approximated  by
$e^{-I_{min}}$,  where $I_{min}$ is the classical
action for the {\it non-extremal} case,
{\it minimized} with respect to $q,~m$. The result,
which should be a function of $\beta,~\Phi$, can be read  off
\cite{york}. It leads to an entropy equal to a quarter of  the
area for all values of $\beta, ~\Phi$.
The averages  $Q, ~M$, as opposed to the parameters $q,~m$, are
calculated from $\beta,~\Phi$. We are interested in $|Q|=M$, {\it
i.e.,} the extremal black hole. This is obtained for limiting values
\bb
\beta\to\infty,~~|\Phi|\to 1,~{\rm with}~\beta(1-|\Phi|)~
{\rm finite}
\ee
for the ensemble parameters and is described by the effective action
\bb
I=\pi M^2={(\beta(1-|\Phi|))^2\over 4\pi}.\ee

It is worth emphasizing again that for extremal black holes,
the parameters $\beta,\Phi$ necessarily enter in the combination
$\gamma\equiv\beta(1-|\Phi|)$.
This combination does occur here as it also does in the case
with purely extremal topology \cite{GMcom}.

Thus in the limit the partition function takes the form
\bb
Z=e^{-{\gamma^2\over 4\pi}}=e^{-\pi M^2}=e^{-{A\over 4}}.\ee
This continues to correspond to an entropy
of a quarter of the area of the horizon, which
is the value of the entropy consistent with the microscopic model.

To reach this goal, we defined extremality {\it not}
by  equating  the  classical parameters $q,m$ before quantization,
but in terms of the
averages  $Q,M$  which  are  calculated  from  the
ensemble  characteristics  $\beta, \Phi$ and which reduce to $q,m$
for the configuration with the minimum action
in the semiclassical approximation.  It  is  because  of  this
altered definition,  and  the use of the sum over  topologies, that
non-extremal configurations have entered and we
have obtained the area law for the entropy instead of the smaller
values  obtained  in  \cite{HHR,GMtwo}.  This  suggests   that   the
microscopic model  discussed above implicitly involves
a quantization procedure where the classical euclidean topology is ignored
and the condition of extremality imposed only after quantization.

It may be clarified here that this need not be
{\it the only correct} way of quantization. In other areas of physics,
there are different, often inequivalent, ways of quantization, {\it
many of them equally acceptable}. The
results referred to in the previous section correspond to quantization
with fixed euclidean topology, while the new models
agree with, but do not explicitly involve, a sum over topologies.

\section*{Acknowledgment}
It is a pleasure to record here the name of my collaborator, Amit Ghosh.
I am grateful to the organizers of the Workshop for the local hospitality
provided at Puri.

\end{document}